\documentstyle[12pt]{article}
\def\be {\begin{equation}}
\def\ee {\end{equation}}
\def\ba {\begin{eqnarray}}
\def\ea {\end{eqnarray}}
\def\nn {\nonumber}
\def\bi {\begin{itemize}}
\def\ei {\end{itemize}}
\begin{document}
\def\bea{\begin{eqnarray}}
\def\eea{\end{eqnarray}}
\title{\bf {Gauge and Gravitational Anomalies and Hawking Radiation of Rotating BTZ Black Holes }}
 \author{M.R. Setare  \footnote{E-mail: rezakord@ipm.ir}
  \\{Department of Science,  Payame Noor University. Bijar. Iran}}
\date{\small{}}

\maketitle
\begin{abstract}
In this paper we obtain the flux of Hawking radiation from Rotating
BTZ black holes from gauge and gravitational anomalies point of
view. Then we show that the gauge and gravitational anomaly in the
BTZ spacetime is cancelled by the total flux of a $2$-dimensional
blackbody at the Hawking temperature of the spacetime.
 \end{abstract}
 \newpage
 \section{Introduction}
It is commonly believed that any valid theory of quantum gravity
 must necessary incorporate the Bekenestein-Hawking definition of
 black hole entropy \cite{bek,haw} into its conceptual framework.
 However, the microscopic origin of this entropy remains an
 enigma. Many efforts have been devoted to this theme. An approach to the Hawking radiation is to calculate the
energy-momentum tensor in the black hole backgrounds. However the
usual expression
 for the stress tensor includes singular products of the field
 operators for stress tensor. Renormalization theory of the stress
 tensor claims to solve this problem, but it must be mentioned that
 the usual scheme of renormalization includes complexity and somewhat
 ambiguity. In semiclassical framework for yielding a sensible theory of back
  reaction Wald \cite{wald} has developed an axiomatic approach.
  There one tries to obtain an expression for the renormalized
  $T_{\mu\nu}$ from the properties (axioms) which it must fulfill.
  The axioms for the renormalized energy momentum tensor are as follow:

  1-For off-diagonal elements standard results should be obtained.

  2-In Minkowski space time standard results should be obtained.

  3-Expectation values of energy momentum are conserved.

  4-Causality holds .

  5-Energy momentum tensor contains no local curvature tensor depending on
  derivatives of the metric higher than second order.

  Two prescriptions that satisfy the first four axioms can differ
  at most by a conserved local curvature term. Wald \cite{wald2}, showed any prescription
  for renormalized $T_{\mu \nu}$ which is consistent with axioms 1-4 must yield the
  given trace up to the addition of the trace of conserved local curvature. It
  must be noted  (that trace anomalies in stress-tensor, that is, the nonvanishing
  $T^\mu _\mu$ for a conformally invariant field after
  renormalization)
  are originated from some quantum behavior \cite{jac}. In two dimensional
  space time one can show that a trace-free stress tensor can not be
  consistent with conservation and causality
  if particle creation occurs \cite{chris}. A trace-free, conserved stress
  tensor in two dimensions must always remain zero if it is
  initially zero. One can show that
it is  possible to use an  trace anomaly  to the energy-momentum
tensors of quantum fields under boundary conditions in a black hole
background \cite{set}, see also \cite{solo} as an example which both
conformal and gravitational anomalies are present.  In this
reference the two-dimensional stress tensor have determined by
requiring that it is regular at the black hole horizon. It is shown
in this paper, that the gravitational anomaly leads to additional
flow in the $(rt)$ component of the stress tensor. In four
dimensions, just as in two dimensions, a trace-free stress tensor
which
  agrees with the formal expression for the matrix elements between orthogonal
  states can not be compatible with both conservation laws and
  causality. Recently,  Robinson and Wilczek suggested
a new derivation of Hawking radiation from Schwarzschild black holes
through gravitational anomalies \cite{Robinson:2005pd}. The anomaly
in field theory occurs if a symmetry of the action or the
corresponding conservation law, valid in the classical theory, is
violated in the quantized version. A gravitational anomaly is an
anomaly in general covariance which shows up as the non-conservation
of the energy-momentum tensor. In Robinson-Wilczek work, the Hawking
radiation is understood as compensating flux to cancel gravitational
anomalies at the horizon
\cite{Robinson:2005pd,Iso:2006wa,Iso:2006ut, soda, elias}. As have
indicated in \cite{Iso:2006wa}, Robinson-Wilczek approach is
effectively equivalent with dilaton-coupled trace anomaly in two
dimensions (to see the conformal anomaly for most general $2D$ dilaton coupled scalar-dilaton system
with arbitrary dilaton couplings refer to  \cite{odi2}). \\
In 1992 Ba\~nados, Teitelboim and Zanelli (BTZ)
\cite{banados1,banados2} showed that $(2+1)$-dimensional gravity has
a black hole
 solution. This black hole is described by two (gravitational) parameters,
the mass $M$ and the angular momentum (spin) $J$. It is locally
AdS and thus it differs from Schwarzschild and Kerr solutions
since it is asymptotically anti-de-Sitter instead of flat
spacetime. Additionally, it has no curvature singularity at the
origin.
 AdS black holes, are members of this two-parametric family
of BTZ black holes and they are very  interesting in the framework
of string theory and black hole physics
\cite{strominger1,strominger2}. The Kaluza-Kelin reduction of $3D$
gravity with minimal scalars to $2D$ dilaton-Maxwell gravity with
dilaton coupled scalars and the rotating BTZ black hole can be also
described by $2D$ charged dilatonic black hole \cite{odi1}.
\\ In this paper we show that
Robinson-Wilczek method  can also be applied to a rotating BTZ black
holes. We show that the gauge and gravitational anomaly in this
 black hole background is cancelled by the total flux of a
$1+1$ dimensional blackbody at the Hawking temperature of this
black hole.
\section{Robinson-Wilczek Approach to the Gravitational
 Anomalies}
 In this section, we will review the gravitational anomaly
method~\cite{Robinson:2005pd,Iso:2006wa,Iso:2006ut, soda, elias}.
Consider the metric of the type,
\begin{equation}\label{metric1}
 ds^2 = -f(r)dt^2 + \frac{1}{f(r)}dr^2 + r^2 d\Omega_{D-2}^2 \ ,
\end{equation}
where $d^{2}\Omega_{(D-2)}$ is the line element of the
(D-2)-dimensional unit sphere and the metric element $f(r)$
depends on the matter distribution. The horizon is located at
$r=r_H$, where $f(r_H)=0$. The scalar field theory on this metric
can be reduced to the 2-dimensional theory. The action of the
scalar field is \bea
 S[\varphi] &=& \frac{1}{2}\int d^D x \sqrt{-g}\,\varphi \nabla^2 \varphi\nn\\
&=& \frac{1}{2}\int d^D x\,r^{D-2} \sqrt{\gamma}  \times \varphi
\left( -\frac{1}{f}\partial_t^2 + \frac{1}{r^{D-2}}\partial_r
r^{D-2} f \partial_r + \frac{1}{r^2}
 \Delta_{\Omega}\right) \varphi   ,
\eea where $\gamma$ is the determinant of $d\Omega_{D-2}^2$ and
$\Delta_\Omega$ is the collection of the angular derivatives. Now
we take the limit $r \rightarrow r_H$ and leave only dominant
terms. Thus, the action becomes \bea
 S[\varphi]& =& \frac{{r_H}^{D-2}}{2}\int d^D x \sqrt{\gamma} \,\varphi \left(
 -\frac{1}{f}\partial_t^2 + \partial_r f \partial_r
  \right) \varphi\nn\\
&=& \sum_n \frac{{r_H}^{D-2}}{2} \int dt dr \,\varphi_n \left(
 -\frac{1}{f}\partial_t^2 + \partial_r f \partial_r
  \right) \varphi_n
\eea in the second line $\varphi$ is expanded by
$(D-2)$-dimensional spherical harmonics. This action is infinite
set of the scalar fields on the 2-dimensional metric
\begin{equation}
 ds^2 = -f(r)dt^2 + \frac{1}{f(r)}dr^2   \ .
\end{equation}
Thus, we can reduce the scalar field theory in $D$-dimensional
black hole spacetime to that in 2-dimensional spacetime near the
horizon. The gravitational anomaly of the chiral scalar field in
$1+1$ dimensions is given as \cite{bert2,louis} \be \nabla_{\mu}
T^{\mu}_{\nu}=\frac{1}{96\pi\sqrt{-g}}\epsilon^{\beta\delta}
\partial_{\delta}\partial_{\alpha}
\Gamma^{\alpha}_{\nu\beta} \hspace{1ex}. \ee The aforesaid
gravitational anomaly is purely timelike and can be given by \be
\nabla_{\mu}T_{\nu}^{\mu}\equiv
A_{\nu}\equiv\frac{1}{\sqrt{-g}}\partial _{\mu}N^{\mu}_{\nu}
\label{anomaly} \ee where the quantities $N^{\mu}_{\nu}$ are
defined as \be
N^{\mu}_{\nu}=\frac{1}{96\pi}\epsilon^{\beta\mu}\partial_\alpha
\Gamma^{\alpha}_{\nu\beta} \ee and the epsilon tensor reads \be
\epsilon^{\mu\nu}= \left(
\begin{array}{cc}
0& 1\\
-1& 0\\
\end{array}
\right) \hspace{1ex}. \ee For the specific Schwarzschild type
black hole spacetime described by (\ref{metric1}), the components
of  $N^{\mu}_{\nu}$ are \bea
N^{t}_{t}&=&N^{r}_{r}=0\nn\\
N^{r}_{t}&=&\frac{1}{192\pi}\left(f'^2 + f''f\right)\\
N^{t}_{r}&=&-\frac{1}{192\pi f^2}\left(f'^2 - f''f\right)\nn
\hspace{1ex}. \eea Therefore the quantity $\Phi$ that describes
the pure flux reads \bea
\Phi&=&N^{r}_{t}\Big|_{r_{H}}\\
&=&\frac{1}{192\pi}f'^{2}(r_{H}) \label{flux1} \hspace{1ex}. \eea
It is well known from thermodynamics that the surface gravity
$\kappa$ of the specific Schwarzschild type black hole is given by
\bea
\kappa&=&\frac{1}{2}\frac{\partial f}{\partial r}\Big|_{r=r_{H}}\\
&=&\frac{1}{2}f'(r_{H}) \label{surface1} \hspace{1ex}. \eea It is
also known from black hole thermodynamics that the Hawking
temperature of this black hole is given as \bea
T_{H}&=&\frac{\kappa}{2\pi}\\
&=&\frac{f'(r_{H})}{4\pi} \eea while a beam of massless black body
radiation moving in the positive radial direction at a temperature
$T_{H}$ in a spacetime  has a flux of the form \be
\Phi=\frac{\pi}{12}T^2 \label{flux2} \hspace{1ex}. \ee Therefore
it is evident that the flux (\ref{flux1}) required to cancel the
gravitational anomaly at the horizon is precisely the thermal flux
from a black hole at the Hawking temperature.
 \section{Rotating BTZ Black Hole}
 The discussion in the previous section
can be extended to the  rotating BTZ black holes.
 The black hole
solutions of Ba\~nados, Teitelboim and Zanelli
\cite{banados1,banados2} in $(2+1)$ spacetime dimensions are
derived from a three dimensional theory of gravity \be S=\int
dx^{3} \sqrt{-g}\,({}^{{\small(3)}} R+2\Lambda) \ee with a
negative cosmological constant ($\Lambda=\frac{1}{l^2}>0$).
\par\noindent
The corresponding line element is \be ds^2 =-\left(-M+
\frac{r^2}{l^2} +\frac{J^2}{4 r^2} \right)dt^2
+\frac{dr^2}{\left(-M+ \displaystyle{\frac{r^2}{l^2} +\frac{J^2}{4
r^2}} \right)} +r^2\left(d\theta -\frac{J}{2r^2}dt\right)^2
\label{metric}\ee whit metric element: \be
f(r)=\left(-M+\frac{r^2}{l^2} +\frac{J^2}{4
r^2}\right)\label{metric2}
 \ee
with $M$ the Arnowitt-Deser-Misner (ADM) mass, $J$ the angular
momentum (spin)
 of the BTZ black hole and $-\infty<t<+\infty$, $0\leq r<+\infty$, $0\leq \theta <2\pi$.
\par \noindent
The outer and inner horizons, i.e. $r_{+}$ (henceforth simply
black hole horizon) and $r_{-}$ respectively, concerning the
positive mass black hole spectrum with spin ($J\neq 0$) of the
line element (\ref{metric}) are given as  \be
r^{2}_{\pm}=\frac{l^2}{2}\left(M\pm\sqrt{M^2 -
\displaystyle{\frac{J^2}{l^2}} }\right) \label{horizon1} \ee and
therefore, in terms of the inner and outer horizons, the black
hole mass and the angular momentum are given, respectively, by \be
M=\frac{r^{2}_{+}}{l^{2}}+\frac{J^{2}}{4r^{2}_{+}}\label{mass}\ee
and \be J=\frac{2\, r_{+}r_{-}}{l}\label{ang}\ee with the
corresponding angular velocity to be \be\Omega=\frac{J}{2
r^{2}}\label{angvel}\hspace{1ex}.\ee
\par\noindent The Hawking temperature $T_H$ of the black hole
horizon is \cite{kumar1} \bea T_H &=&\frac{1}{2\pi
r_{+}}\sqrt{\left(\displaystyle{\frac{
r_{+}^2}{l^2}+\frac{J^2}{4r_{+}^2}}\right)^2-\displaystyle{\frac{J^2}{l^2}}}\nn\\
&=&\frac{1}{2\pi r_{+}}\left(\displaystyle{\frac{
r_{+}^2}{l^2}-\frac{J^2}{4r_{+}^2}}\right)\label{temp1}
\hspace{1ex}.\eea \par\noindent If one neglect classically
irrelevant ingoing modes near the horizon, the effective
two-dimensional theory becomes chiral near the horizon and the
gauge symmetry or general coordinate covariance becomes anomalous
due to the gauge or gravitational anomalies. The current is
conserved \be \partial_{r}J_{(o)}^{r}=0 \label{con1} \ee outside
the horizon. In the region near the horizon, since there are only
outgoing fields, the current satisfies the anomalous equation \be
\partial_{r}J_{H}^{r}=\frac{m^2}{4\pi}\partial_{r}A_t \label{con2} \ee
where $m$ is the $U(1)$ charge of two-dimensional massless field.
One can solve these equations in each regions as \be J_{(o)}^{r}=C_0
\label{con3} \ee \be J_{H}^{r}=C_H+\frac{m^2}{4\pi}(A_t(r)-A_t(r_+))
\label{con4} \ee where $C_0$ and $C_H$ are integration constants.
Under gauge transformations, variation of the effective action is
given by \be -\delta W=\int d^{2}x \sqrt{-g_{2}}\lambda
\nabla_{\mu}J^{\mu} \label{eff1} \ee where $\lambda$ is a gauge
parameter, and \be J^{\mu}=J_{(o)}^{\mu}\theta_{+}(r)+
J_{H}^{\mu}H(r) \label{con5}\ee here
$\theta_{+}(r)=\theta(r-r_+-\epsilon)$ and $H(r)=1-\theta_{+}(r)$.
By using the anomaly equation we have \be -\delta W=\int d^{2}x
\lambda
[\delta(r-r_+-\epsilon)(J_{(0)}^{r}-J_{(H)}^{r}+\frac{m^2}{4\pi}A_t)+\partial_{r}(\frac{m^2}{4\pi}A_tH)].
\label{eff2} \ee The total effective action must be gauge invariant
and the last term should be cancelled by quantum effects of the
classically irrelevant ingoing modes. The quantum effect to cancel
this term induced by the ingoing modes near the horizon. The
coefficient of the delta-function should also vanish, which relates
the coefficient of the current in two region, \be
C_0=C_H-\frac{m^2}{4\pi}A_t(r_+) \label{ceq}\ee $C_H$ is the value
of the consistent current at the horizon. In order to determine the
current flow, we need to fix the value of the current at the
horizon. Since the covariant current is written as
$\tilde{J^{r}}=J^{r}+\frac{m^2}{4\pi}A_tH$, the condition
$\tilde{J^{r}}(r_+)=0$ determines the value of the charge flux to be
\be
C_0=-\frac{m^2}{2\pi}A_t(r_+)=-\frac{m^2}{2\pi}\Omega(r_+)=-\frac{m^2}{2\pi}\frac{J}{2r_{+}^{2}}=
-\frac{m^2 r_{-}}{2\pi l r_+} \label{ano}\ee Similarly one can
determine the flux of the stress tensor radiated from rotating BTZ
black holes. The total flux of the energy-momentum tensor is given
by \be a_0=
\frac{m^2}{4\pi}A_{t}^{2}(r_+)+N_{t}^{r}(r_+)=\frac{m^2}{4\pi}\Omega^{2}(r_+)+\frac{1}{192\pi}
f'^2 = \frac{m^2 r_{-}^{2}}{4\pi l^2 r_{+}^{2}}+\frac{1}{192\pi}f'^2
\label{graan}\ee Now we evaluate the quantity $f'$ on the horizon.
For the rotating BTZ metric this quantity is \be
f'|_{r_{+}}=\frac{2r_{+}}{l^2}-\frac{J^2}{2r_{+}^{3}}
\label{quantityN} \ee By comparing with the Hawking temperature
$T_H$ of the black hole horizon we see \be f'|_{r_{+}}=4\pi T_H
\label{comp}  \ee Then we can rewrite eq.(\ref {graan}) as \be a_0=
\frac{m^2 r_{-}^{2}}{4\pi l^2 r_{+}^{2}}+\frac{\pi}{12}T^{2}_{H}
\label{flux} \ee This value of the flux is the same as the Hawking
flux from rotating BTZ black holes.
\section{conclusion}
In this paper we have considered a quantum field in a rotating BTZ
black hole background. Near the horizon, the field can be
effectively described by an infinite collection of (1+1)-dimensional
fields. In the case of rotating BTZ black hole, the metric is
azimuthal symmetric and the  angular momentum is conserved. Because
of this isometry, the effective two-dimensional theory for each
partial wave has $U(1)$ gauge  symmetry. The effective background
gauge potential for this $U(1)$ symmetry is written in terms of the
metric while the quantum field in the rotating BTZ background has a
charge $m$ of this gauge symmetry, where $m$ is an angular quantum
number. The effective theory is now interpreted as that of charged
particles in a charged black hole in $d=2$ \cite{Iso:2006wa} (see
also \cite{Iso:2006ut}). Then we have shown that the Hawking
radiation from rotating BTZ blacks holes can be obtain from gauge
and gravitational anomalies. We have shown that the gauge and
gravitational anomaly that appears in the BTZ spacetime  is
cancelled by the total flux of a $1+1$ dimensional blackbody
radiation at the Hawking temperature.

\end{document}